# The Many Publics of Science:
# Using Altmetrics to Identify Common Communication Channels by Scientific field

Daniel Torres-Salinas[1], Domingo Docampo[2], Wenceslao Arroyo-Machado[1], and Nicolas Robinson-Garcia[1]

[1] Department of Information and Communication Sciences, University of Granada

[2] Atlantic Research Center for Information and Communication Technologies, University of Vigo

## Abstract

Altmetrics have led to new quantitative studies of science through social media interactions. However, there are no models of science communication that respond to the multiplicity of non-academic channels. Using the 3653 authors with the highest volume of altmetrics mentions from the main channels (Twitter, News, Facebook, Wikipedia, Blog, Policy documents, and Peer reviews) to their publications (2016-2020), it has been analyzed where the audiences of each discipline are located. The results evidence the generalities and specificities of these new communication models and the differences between areas. These findings are useful for the development of science communication policies and strategies.

# 1 Introduction

The rise of social media on the one hand, and open science on the other, have increased the complexity of the science communication process. The number of platforms has multiplied, while the directionality of the communication process has expanded, with multiple stakeholders at the two ends of the process, and not always involving science communicators (Bray et al., 2012). Scientific literature is now widely discussed and shared on platforms such as Twitter, Facebook, Wikipedia or blogs (Arroyo-Machado et al., 2022; Alperin et al., 2019; Enkhbayar et al., 2020; Robinson-Garcia et al., 2017; Shema et al., 2012), along with traditional news media and online mass media (Moorhead et al., 2021). Furthermore, scientists are being pushed through Open Science policies to increase the transparency of their processes and actively engage with society (Grand et al., 2012; Zastrow, 2020). While transparency and openness are desirable in principle, they also open the door to publicly discussing scientific findings which remain to be proven. This was the case, for instance during the COVID-19 pandemic, when preprints were massively made available (Fraser et al., 2021; T. Nane et al., 2021), stirring online public discussions based on questionable data (Torres-Salinas et al., 2020).

The massive and multi-channelled consumption of scientific literature presented many challenges as to how to mitigate issues related to misinformation or mistrust in science, and how scientific knowledge transfers to society potentially leading to informed decisions, especially in areas such as health policy (van Schalkwyk et al., 2020) or risk management (Nane et al., 2021) among many others. Scientific literature in this regard has often focused on topics such as decision-making processes, identification of audiences, reporting and communication strategies or the identification of the appropriate channels for communication (Fischhoff, 2013; Logan, 2001). In this paper, we focus on the latter aspect.

Our goal is to identify the channels by which individuals share, discuss, and engage with scientific literature. More specifically, we aim at looking at differences by scientific field in order to understand how field-specific dynamics affect the channels by which scientific publications are shared. For this, we propose the use of scientometric techniques for identifying non-academic mentions to scientific literature, also known as 'altmetrics' (Priem, 2014). We look into a set of over 200,000 researchers and identify their scientific outputs. Then we select those who accrue the largest number of altmetric mentions to their oeuvre and analyse the channels by which their outputs are shared in a variety of traditional and online channels (i.e., Twitter, blogs, Facebook, policy documents, news media). Individuals are assigned to fields based on their scientific output. We then characterize each field's singularities based on the communication channels that are specifically used.

The contribution of this paper is twofold. First, it integrates two streams of literature: 1) empirical studies on altmetrics and its potential use in quantitative studies of science, and 2) theoretical studies on science communication models. Second, by identifying the importance of communication channels by scientific field, it provides empirical evidence on which of these channels should be explored in order to better understand how science is being discussed and consumed by laypeople, and hence, inform decision makers on the design of communication campaigns and strategies at the institutional level.

The paper is organized as follows: first, we review the literature and focus on the science-society interactions produced in social media and the multiplicity of communication channels, as well as on the particularities by scientific fields. Secondly, we describe our data and the statistical analysis performed. Third, we present the results of our analysis, showing the singularities detected in the different scientific fields. Finally, we conclude by discussing our findings and highlighting their usefulness for scientific communication.

## 2 Literature review

### 2.1 Science-society interactions and social media

Although science communication has been historically modelled as unidirectional (Riise, 2008), we know that science and society interact through a wide range of mechanisms, influencing and shaping each other's development (Ramos-Vielba et al., 2022). These mechanisms are non-linear, nor they are homogeneous across different types of knowledge, stakeholders, or platforms. Neither there is an ascendency of science over society or vice versa (Camic et al., 2011). Science communication takes place throughout different actors in which the participation of the specialist or expert is no longer needed (Logan, 2001). Social media maximise this type of non-linear communication model, by providing users with the platform for both: delivering and sending messages publicly (Costas et al., 2021). Actors of all kinds can potentially participate in this dialogue (Davies & Hara, 2017; Kavanaugh et al., 2012; Robinson-Garcia et al., 2018).

Studies on science consumption show that open science has fostered the use of research for decision-making, professional development, educational purposes or even for leisure (Hicks et al., 2022). Journalists are still key players in the dissemination of research and serve as translators and brokers of science for the wider public (Brechman et al., 2009; van Schalkwyk & Dudek, 2022). Evidence suggests that news media articles introducing novel research to the wider public can greatly impact professional practice for instance (Hicks et al., 2019). This role of the media is further explored by Isett & Hicks (2020), who revise through a series of examples, how news media reinforce policy making through their filtering and use of scientific literature. With regard to social media, there is a wide number of studies focused on the dissemination of scientific information through different social media platforms. Most of these studies are focused on the phenomenon of disinformation. For instance, related to the use of Twitter during the COVID-19 pandemic (Bogomoletc et al., 2021), or in relation with the anti-vaccination movement (van Schalkwyk et al., 2020), or the spread of fake news throughout WhatsApp messaging (Elías & Catalan-Matamoros, 2020) among others.

These examples show that the heterogeneity of channels, sources of information and audiences is increasing the complexity of the science communication process and of those interested in designing effective strategies of communication (Fischhoff, 2013). Finding approaches by which such communication can be identified and analyzed is a crucial step to better monitoring and designing communication programs.

## 2.2 Monitoring the social attention of science with altmetrics

The field of evaluative scientometrics has been exploring for over more than a decade, how scientific literature is consumed beyond academia, that is, beyond the use of citations. In 2010, the term 'altmetrics' was coined (Priem, 2014). It was the result of three major drivers (Moed, 2016). First, the pressure from funding agencies to quantify and measure the societal impact of research (Spaapen & Drooge, 2011). Second, the rapid digital transformation of the scholarly communication system along with the rise of social media (Sugimoto et al., 2017). Third, the Open Access movement which calls for greater transparency and accessibility to scientific literature (Chan et al., 2002). At first, altmetrics were seen as an alternative to citation metrics due to their capacity to capture mentions in social media at a greater speed than citations go beyond the academic circuit (Eysenbach, 2011; Wouters & Costas, 2012). But their loose definition and a lack of evidence of their use as substitutes for citations soon overcame their prospects as an alternative to citation metrics (Bornmann, 2014; Haustein, 2016).

Altmetrics originally referred to 'the study of new metrics based on the Social Web for analysing, and informing scholarship' (Priem et al., 2010). However, since they were first defined, the number of metrics included under such term has increased and gone beyond social media. This is due to the essential role commercial altmetric data providers have played on the expansion and use of altmetrics, namely the two main scientific publishers Springer Nature and Elsevier, through their subsidiaries Altmetric.com and PlumX, respectively (Zahedi & Costas, 2018). Currently, altmetrics refer to mentions in social media as well as citations in policy documents and clinical guidelines, presence in library catalogs or mentions in news media, and the number continues increasing as altmetric providers expand the battery of indicators they offer.

From a science communication point of view, the role played by these outlets is key to understanding how science is currently perceived by laypeople. In this regard, altmetrics have proven to be an important source of information for mapping social engagement of scientists with the public (Robinson-Garcia et al., 2018), identifying communities of interest (Díaz-Faes et al., 2019), identifying socially relevant topics (Haunschild et al., 2019; Robinson-Garcia et al., 2019), or analysing controversy and debate around science (G. F. Nane et al., 2021; van Schalkwyk et al., 2020). This has happened by abandoning raw counts and adopting more advanced techniques derived from social network analysis (Arroyo-Machado et al., 2021; Hellsten et al., 2019) and science mapping (Costas et al., 2021).

Some attempts at using altmetric data have already been made in the field of science communication. For instance, Yeo et al. (2017) complement a case study on a controversial debate around a scientific paper published in Science in the field of genetics with altmetrics. Alperin et al. (2019) use altmetrics to study diffusion patterns on Twitter of scientific literature. Finally, van Schalkwyk et al. (2020) tracked altmetric mentions to COVID-19 preprints to study their reporting in South African news outlets. Still, an overall analysis on field-specific differences of the most common channels used is still lacking.

## 2.3 Differences between communication channels using altmetrics

Still, a general picture of the role these altmetric sources play as communication channels by fields is lacking. We know that there are fundamental differences between altmetrics sources. Hence, many of the sources offered by altmetric data providers are normally discarded (Robinson-García et al., 2014) due to their low incidence and uncertain definition (e.g., Google Plus, YouTube, Syllabi), while others are studied in more detail (e.g., Twitter, news media, Wikipedia mentions, policy documents). For instance, studies on Twitter point out at its relevance in identifying topics of interest or communities of shared interest (Haunschild et al., 2019; van Schalkwyk et al., 2020), but it is not a good source to find public engagement with science (Fang et al., 2021; Robinson-Garcia et al., 2017). In the case of Facebook, altmetric providers usually underestimate the volume of research shared, as most of it is done privately and not in public pages and groups (Enkhbayar et al., 2020).

Studies on Wikipedia indicate the importance of learning how this widely used source feeds from scientific literature (Singh et al., 2021), and show how science is organized from the point of view of the public as opposed to the academic perspective (Arroyo-Machado et al., 2020). News media and blogs have been found to be the easiest proxy to use when trying to understand how science is presented to the public. However, concerns have been raised as to how these data are collected by altmetric providers (Ortega, 2020). Fleerackers et al., (2022) indicate that, despite having high precision and reasonable recall overall, Altmetric.com is less reliable at identifying mentions to scientific literature when no bibliometric information is provided, and affecting specially monographs, conference presentations, dissertations, and non-English literature. Finally, a very interesting case is that of policy documents. Despite their low coverage (Bornmann et al., 2016; Fang, Costas, et al., 2020), policy documents have received lots of attention from scientometricians. This channel of communication allows tracking the translation of science into policy practices and recommendations. This has led many to the launching of a new data source specifically focused on capturing mentions to scientific literature in policy documents (Fang, Dudek, et al., 2020; Pinheiro et al., 2021).

## 2.4 Differences between scientific fields

Studies analyzing disciplinary differences regarding altmetrics seem to be somehow contradictory (Sugimoto et al., 2017, p. 2046). In many cases, this has to do with differences on the altmetric provider used (Ortega, 2018; Zahedi & Costas, 2018). In other cases, this has to do with the scientific database used to identify science literature. For instance, Fang, Costas, et al. (2020) search for mentions in Altmetric.com to papers indexed in the Web of Science. They find that Computer Science and Mathematics are the less discussed fields. On the contrary, Hassan et al. (2017) use Scopus and find that it is literature from Chemistry and Materials Science the one receiving less social attention.

Disciplinary differences are influenced by a language bias towards the English language (Ortega, 2018) and geographic areas (Alperin, 2015). The use or lack of use of certain communication channels by countries is also important. Lack of access to Twitter in countries such as China or Iran will affect globally how much this channel is used in certain fields to discuss scientific literature. Also, while Social Sciences and Humanities tend to be highly discussed in social media, there is evidence that when looking at non-English science, this is not the case (Torres-Salinas et al., 2016).

Overall, there seems to be consensus that the most discussed science is that related to the fields of Biomedicine and Health Sciences (Fang, Costas, et al., 2020; Hassan et al., 2017; Zahedi et al., 2014). However, different fields will have their own preferred communication channels. For instance, Zahedi et al. (2014) show that the Law, Arts & Humanities tend to be highly mentioned in Wikipedia when compared with the rest of the fields. Still, Arroyo-Machado et al. (2020) report that it is the fields of Medicine, Biochemistry, and Genetics and Molecular Biology the ones which accumulate the highest number of mentions in Wikipedia.

Also, different communication channels will be used depending on the specific topic within a discipline. Robinson-Garcia et al. (2019) showed how, in the case of Microbiology, different topics would be highlighted depending on the communication channel used (in their case, Twitter, news media and policy documents). This point was also stressed by Fang, Costas, et al., (2020). The importance of understanding, not only which disciplines raise more social interest from a communication perspective, but also which is the desired communication channel for each discipline and topic is essential to be able to design and effectively monitor science communication programmes. Different communication channels will show different dimensions or aspects of how science connects with society, offering a partial perspective on how science is perceived (Noyons, 2019). Hence there is the risk of focusing on the wrong communication channel when trying to assess the effectiveness of a science communication plan.

# 3 Material and methods

Data retrieval was performed on March 3, 2021. We first downloaded all bibliographic records published between 2016 and 2020 for which an author with Spanish affiliation was listed from Web of Science, and retrieved only the following document types: articles, editorial material, letters and proceedings. We considered for following citation indexes: Science Citation Index Expanded (SCI-Expanded), Social Sciences Citation Index (SSCI), Arts & Humanities Citation Index (A&HCI), and Emerging Sources Citation Index (ESCI). 434,827 records were obtained and exported to InCites to reclassify records categorized as "Multidisciplinary" in Web of Science. Altmetric mentions for these publications were retrieved from the Altmetric Explorer. This process was carried out by querying the database through the digital object identifier (DOI) of the publications retrieved from Web of Science. 406,621 records included a DOI (93.51%), out of which 238,508 were indexed in Altmetric.com (54.85% of the total) and had a total of 3,596,296 mentions. The records were classified into 22 research fields, which are formed by the 21 included in the Essential Science Indicators (ESI) of Clarivate plus Arts and Humanities. For this purpose, the 254 subject categories of Web of Science were matched with the ESI classification following the equivalence scheme proposed by Arroyo-Machado (2021), which is based on that of Tan (2020).

**Fig. 1** Methodological design from data retrieval to statistical analyses

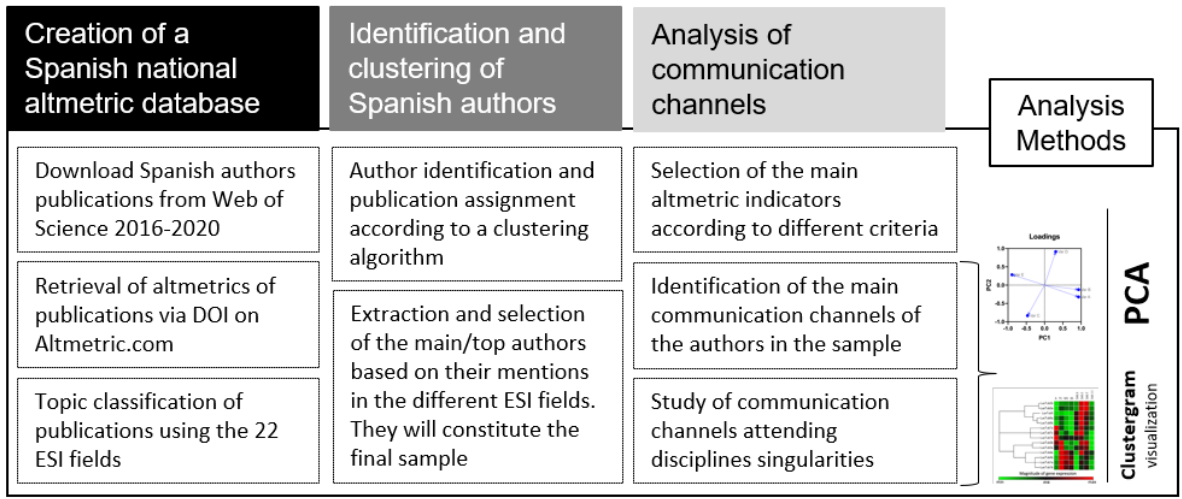

We then identified and disambiguated author names of researchers affiliated to Spanish institutions using the algorithm proposed by Caron and van Eck (2014). We identified a total of 201,891 researchers (Table 1). As a filtering criterion to remove transient authors as well as potential errors of the algorithm, we include only those who have at least three publications indexed in Altmetric.com. For each scientific field, we sorted the set of authors based on their cumulative value of the "Altmetric Attention Score"[1]. We then computed the square root multiplied by three for the total number of authors in each field. This allows us establishing the number of researchers that should be selected in each of the ESI scientific fields. This procedure is used as a standardization process in other sources such as ARWU (Docampo, 2013). The last column of Table 1 indicates the size of the sample of authors used.

**Table 1** Identification of top authors and calculation of the final sample

| | | Step 1 | Step 2 | Step 3 | Step 4 |
|---|---|---|---|---|---|
| | | Total authors identified | Authors with ≥ 3 articles | Total authors after normalization | Top authors included in the study |
| Global – all authors → | GLO | 201891 | 93559 | 918 | 918 |
| | | | | | |
| Essential Science Indicators Field ↓ | | | | | |
| Agricultural Sciences | AGR | 17749 | 6841 | 248 | 250 |
| Arts & Humanities | ART | 2745 | 642 | 76 | 75 |
| Biology & Biochemistry | BIO | 27764 | 7732 | 264 | 250 |
| Chemistry | CHE | 28118 | 12059 | 329 | 325 |
| Clinical Medicine | CLI | 86574 | 28077 | 503 | 500 |
| Computer Science | COM | 9854 | 3926 | 188 | 200 |
| Economics & Business | ECO | 5079 | 2089 | 137 | 150 |
| Engineering | ENG | 29105 | 13490 | 348 | 350 |
| Environment/Ecology | ENV | 24187 | 8571 | 278 | 275 |
| Geosciences | GEO | 11759 | 4740 | 207 | 200 |
| Immunology | IMM | 17104 | 4483 | 201 | 200 |
| Materials Science | MSE | 13756 | 6296 | 238 | 250 |
| Mathematics | MAT | 3796 | 1657 | 122 | 125 |
| Microbiology | MIC | 18736 | 5440 | 221 | 225 |
| Molecular Biology & Genetics | MOL | 18857 | 4677 | 205 | 200 |
| Neuroscience & Behaviour | NEU | 18952 | 5648 | 225 | 225 |
| Pharmacology & Toxicology | PHA | 17228 | 4443 | 200 | 200 |
| Physics | PHY | 13166 | 5617 | 225 | 225 |
| Plant & Animal Science | PLA | 18446 | 7202 | 255 | 250 |
| Psychiatry/Psychology | PSY | 13421 | 4112 | 192 | 200 |
| Social Sciences, General | SOC | 24598 | 6970 | 250 | 250 |
| Space Sciences | SPA | 2450 | 1192 | 104 | 100 |
| Final total authors included → | | | | | 3653 |

---

[1] https://help.altmetric.com/support/solutions/articles/6000233311-how-is-the-altmetric-attention-score-calculated-

For each author, thirteen altmetric indicators were used based on the cumulative number of mentions received by their publication record. We did not consider all of the indicators offered by Altmetric.com, as many of them have serious limitations or biases. Those derived from platforms with an irrelevant number of mentions (e.g., YouTube), platforms with a strong geographic component (e.g., Weibo), platforms that no longer exist or operate (e.g., Google Plus or Syllabi), or including non-verifiable data to ensure its reproducibility (e.g., Mendeley) were removed. The six sources selected for the study are:

- Twitter Mentions: Total number of times an article was mentioned in tweets, including quoted tweets and retweets only.
- Wikipedia Citations. Citations received from Wikipedia entries.
- News Mentions: Total number of times articles are mentioned in news and mainstream media. According to Altmetric.com this list currently extends to about 3,000 English and non-English global news outlets.
- Blogs Mentions: Total number of times papers are mentioned in blogs. Altmetric.com maintains a curated list of over 15,000 academic and non-academic blogs.
- Peer: Nr of received post-publication review in forums such as PubPeer or Publons.
- Facebook: Nr of mentions in Facebook. This is restricted only to public pages and not individual timelines or private groups.

Finally, we ran a Principal Components Analysis (PCA) analysis based on the correlation matrix in SPSS for each ESI area and selected the two main components. We noted for each field the loadings of the indicators in the first and second components. The methodological process is summarised in Fig. 1.

## 4 Results

### 4.1 General Analysis

Table 2 includes the total mentions collected and the mentions received exclusively by the top authors analysed, a total of 3,653 (2% of the authors in the database). These authors have published a total of 63,946 of the scientific papers included in the database. This means that they account for 26% of the publications indexed in Altmetric.com from our original dataset. Moreover, this 2% concentrates a significant part of the mentions. In the case of Twitter, they account for 86% of the 3,183,505 mentions collected. The other cases are not as extreme, although they are significant. For instance, news and Facebook mentions account for 42% of the total. Table 2 shows the overall volume of the different altmetrics calculated. Twitter is the platform that concentrates the most mentions, followed by news mentions (200,772) and Facebook mentions (98,840), while policy mentions and peer review reports only account for 7,785 and 2,303 records respectively.

**Table 2** Total mentions collected in this study and mentions received for papers published for top Spanish authors according to Altmetric.com

|  | Total Mentions for articles published by all authors (=201891) | Total Mentions for articles published by Top Authors (= 3653) | % Mentions for articles published by Top Authors (= 3653) |
|---|---|---|---|
| Twitter | 3183505 | 2722071 | 86% |
| News | 200772 | 85193 | 42% |
| Facebook | 98840 | 41207 | 42% |
| Wikipedia | 11151 | 3762 | 34% |
| Blog | 38729 | 15205 | 39% |
| Policy | 7785 | 2574 | 33% |
| Peer | 2303 | 608 | 26% |

As observed, the Peer indicator seems to be quite irrelevant given its size (only 608 observations for the authors analysed). A preliminary test of the PCA further urged us to remove it from subsequent analyses[2]. The results of the PCA are shown in Table 3. Considering the Rotation Sums of Squared Loadings, the first two components together explain a significant chunk of the variance, more than 70%, a statistically significant percentage (Table 3.1). The results of the analysis also show that all the variables present relevant values in their Communalities (Table 3.3). We can gather from Table 3.1 that Twitter and Facebook compute mostly in component 1, with the largest loadings, 0.767 and 0.758 respectively. Blogs and News Mentions show some degree of similarity and compute in both components with similar loadings. Finally, Wikipedia computes exclusively in Component 2 (0.850), and Policy Mentions exclusively in Component 1 (0.749).

---

[2] Our first PCA showed a Communality of 0.119 for this indicator, hence it was removed.

**Table 3** Principal Component Analysis for global author's sample: variance explained, component matrix & communalities

### 3.1. Total Variance Explained

| | Initial Eigen values | | | Rotation Sums of Squared Loadings | | |
|---|---|---|---|---|---|---|
| | Total | % Of Variance | Cumulative % | Total | % of Variance | Cumulative % |
| Component 1 | 3,311 | 55,179 | 55,179 | **2,410** | 40,165 | **40,165** |
| Component 2 | 0,946 | 15,765 | 70,944 | **1,847** | 30,779 | **70,944** |
| Component 3 | 0,746 | 12,433 | 83,377 | | | |
| Component 4 | 0,457 | 7,609 | 90,985 | | | |
| Component 5 | 0,333 | 5,546 | 96,532 | | | |
| Component 6 | 0,208 | 3,468 | 100,000 | | | |

### 3.2. Rotated Component Matrix

| | Component 1 | Component 2 |
|---|---|---|
| Twitter | 0,767 | 0,347 |
| Wikipedia | -0,009 | 0,850 |
| News Mentions | 0,690 | 0,544 |
| Policy | 0,749 | -0,128 |
| Blogs | 0,460 | 0,755 |
| Facebook | 0,758 | 0,348 |

### 3.3. Communalities

| | Value |
|---|---|
| Twitter | 0,708 |
| Wikipedia | 0,723 |
| News Mentions | 0,772 |
| Policy | 0,577 |
| Blogs | 0,782 |
| Facebook | 0,696 |

The graphical/biplot representation of the principal component analysis is shown in Fig. 4 of the Appendix. Biplots are powerful visualizations for representing multivariate data, and especially useful in scientometric studies (Torres-Salinas et al., 2013). Here we observe that social media plays a central role, blogs and news mentions play an intermediate role, and Wikipedia and policy mentions are positioned in the periphery. Fig. 2 shows how the different channels are combined in disseminating researchers' outputs. We observed that channels are grouped into three clusters. This is consistent with the results obtained via the PCA. That is, one formed by blog and news mentions, a second one formed by Wikipedia and policy mentions, and a third one with Twitter and Wikipedia mentions.

**Fig. 2** Cluster gram at the author level for our sample according to their channels of scientific influence with indications about de predominant ESI field in each cluster

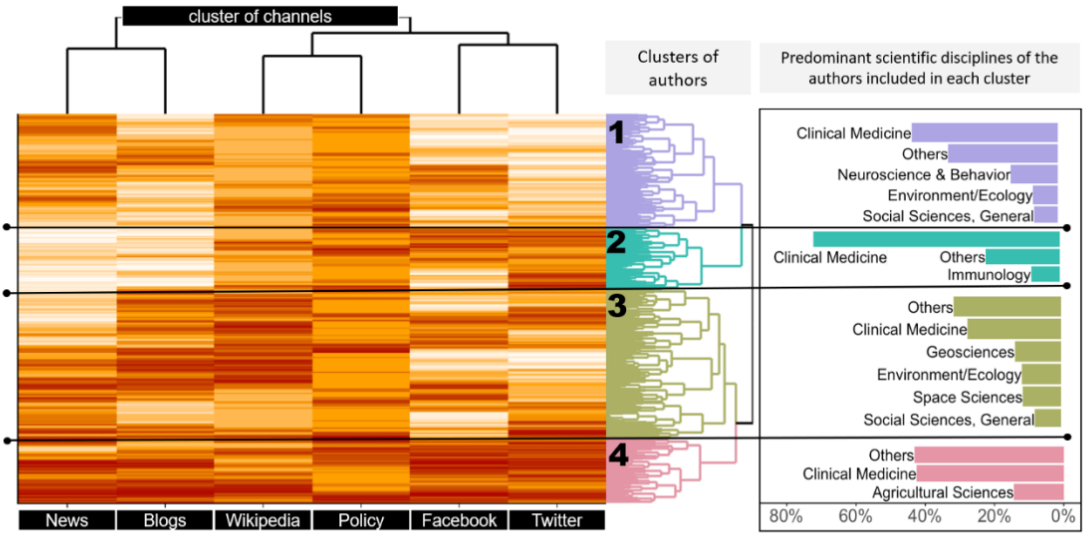

In the case of authors, these are clustered into four groups. Cluster 1 is dominated by authors exhibiting a low presence social media, especially Twitter. In the rest of the platforms they maintain moderate values, although some authors occasionally stand out in their number of mentions in news and policy reports. Cluster 2 groups researchers who stand out mainly on Twitter and have little presence in other media such as news or blogs. Cluster 3 is the cluster with the largest number of authors and subgroups. It is a heterogeneous and miscellaneous cluster where any combination of platforms can occur. Finally, cluster 4 shows a homogeneous composition and an evident pattern. This cluster includes those researchers whose outputs are more widely spread since they manage to reach all audiences. On the right side of the figure, we provide the field composition of each cluster according the ESI classification. In all clusters, Clinical Medicine dominates, especially in cluster 2. Clusters 1 and 3 are more heterogeneous, including fields such as Social Sciences and Environment / Ecology. Cluster 4 integrates a larger number of authors from different areas.

## 4.2 Disciplines singularities

Fig. 3 shows the comparison of the use of communication channels in the 23 ESI fields. For this purpose, we have created a composite cluster gram. Fig. 3 is composed of a bar chart with the sample of authors included in each discipline, a cluster gram classifying the disciplines according to the total number of mentions and a second cluster gram with the median number of mentions in each platform. Disciplines were grouped into four clusters. Next, we describe each of them.

- Cluster 1. It is composed by six scientific disciplines from the life and natural sciences. They show high values in all channels except for policy reports. They stand out in Wikipedia, especially in the case of Geosciences and Space Science, which also have high values in Blogs and News.

- Cluster 2. This group is formed by four scientific fields all of which reach high altmetric values in all channels. Contrarily to cluster 1, it reaches a high number of policy mentions, and less Wikipedia mentions. These fields are Clinical Medicine, Environment / Ecology (both reaching the highest values in Twitter, Facebook and News), Social Sciences and Agricultural Sciences.

- Cluster 3. Formed by eight scientific fields with moderate values, two clearly defined subclusters. The first one includes Health related fields. Although quite homogeneous, some fields stand out, such as Immunology in the case of policy mentions, Microbiology in the case of Facebook or Psychiatry in the case of Wikipedia. The second subcluster is integrated by fields related to Engineering and Exact Sciences. Physics stands out due to the high number of mentions it accrues in blogs and news media. Chemistry and Engineering maintain moderate-low values in all channels.

- Cluster 4. It consists of four scientific fields with the lowest dissemination in most of the channels analyzed. The fields of Mathematics, Computer Science and Arts/Humanities stand out for their low performance. We find a singularity in the field of Economics, which accumulates a high number of policy mentions and a moderate value in blogs mentions.

**Fig. 3** Cluster grams at author level for 22 ESI fields and six channels of the scientific influence with indications about the size of the sample in each field

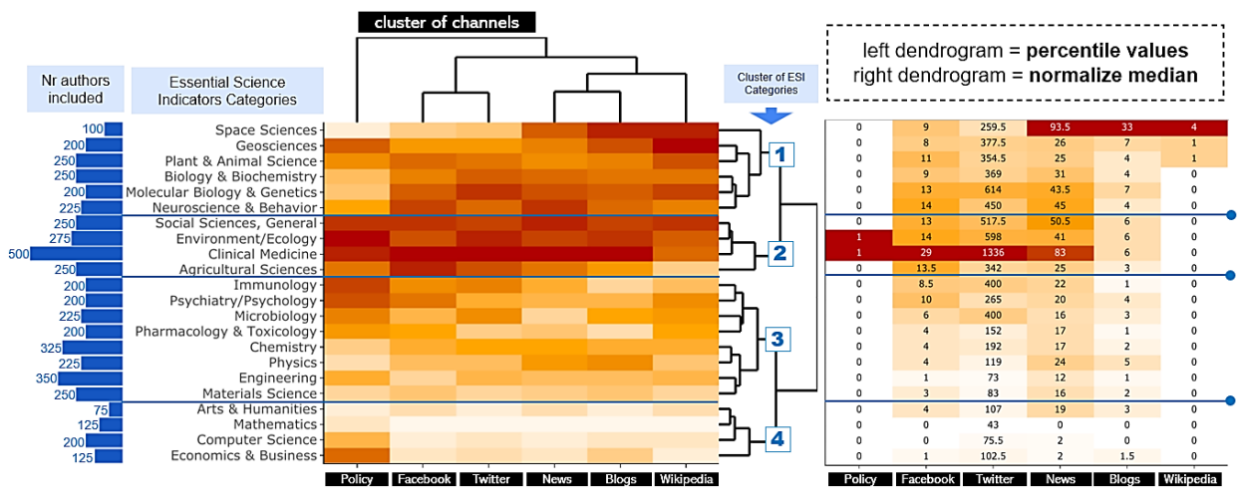

To complement this analysis, we computed a PCA for the 23 ESI fields. The results are provided in Table 4 in the Appendix. In the case of life and health related fields, we observe how the first component is characterized by high loadings, except for policy and Wikipedia mentions, which are represented in the second component. In fields related to Arts, Mathematics and Social Sciences we observe lower loadings of all indicators in both components.

## 5 Discussion and conclusions

This paper aimed at understanding the different paths by which science is consumed by non-academic audiences. Our goal was to analyse differences by scientific field and identify which were the most common channels, as well as combinations of channels. For this, we retrieved a total of 3,596,296 mentions in different non-academic channels directed to 238,508 scientific publications. These publications are authored by researchers affiliated to Spanish institutions between 2016 and 2020. One of the novelties of this paper is that the analyses were conducted at the author level rather than at the publication level. In this way, we classified researchers based on their publication record by their scientific field and retrieve only authors whose oeuvre has been extensively disseminated in non-academic outlets. The second novelty is how altmetric indicators were used. Rather than aiming at looking at societal or broader impact, we use altmetric indicators as a proxy to determine the main channels of scientific communication and, ultimately, where the audiences of scientific publications are. Six channels were studied: Twitter, Facebook, Wikipedia, news media, blogs and policy documents.

First, we conduct a Principal Component Analysis which helps us determine global differences between the selected channels of information. We can distinguish general audiences through social networks such as Twitter and Facebook, which are closely related. On the other hand, we also observe a series of communication channels selecting content and disseminating it to professional (blogs) and generic (news media) audiences. Likewise, there are other ways to reach audiences and that implies the effective use of the results, such as in the educational sphere (Wikipedia) or in the political and decision-making sphere (policy documents). These platforms have smaller audiences as indicated by a lower number of mentions. These platforms also indicate higher levels of engagement with the scientific content. In the case of Wikipedia and policy reports, articles have been read, intellectually integrated and incorporated into a list of bibliographic references.

The combination of these channels generates different scientific communication profiles. The results suggest that there is no standard or archetypal profile since the use of a channel or reaching a certain audience is determined by the scientific field. There are significant differences in the "intensity" of the reception of papers between the scientific fields. We identified fields that accumulate a large part of the altmetrics mentions and compute with high values in all platforms. This the case of Social Sciences, Environment/Ecology or Clinical Medicine, empirical results indicate that these fields communicate scientific results more effectively. We also found areas with difficulties in reaching audiences, especially those related to engineering or exact sciences, a clear example are Mathematics or Arts & Humanities. In the latter case this may be because we are working with a Spanish dataset and hence, there are language and geographical biases in place (Alperin, 2015; van Leeuwen et al., 2001).

Second, we conduct a clustering analysis by field which is complemented with a PCA by field. We do this to better understand disciplinary singularities which may be hidden in the global analysis. Here we find some factors that may explain our results. Some fields like the case of Mathematics or Engineering may simply be more difficult to translate to general audiences. Also, some of them do not have an immediate applicability. In other cases, like the Arts and Humanities, may already have different communication mechanisms and use different outlets other than scientific publications to reach to non-academic audiences (Nederhof, 2006; Arroyo-Machado & Robinson-Garcia, 2021). Regarding differences in the communication channels in which science disseminates, we note the use of Wikipedia by some of the natural sciences, such as Space Science or Geosciences. Another example of this type of singularity is represented by the citations received by economics researchers from policy reports.

Through the use of altmetrics we can quantify the effectiveness of scientific communication, observing significant differences between fields. The results can be useful to guide researchers and managers on identifying the most appropriate channels to disseminate research or developing communication policies and strategies adapted to specific fields and audiences. The requirements for defining a communication policy cannot be the same in areas such as Clinical Medicine, which receives great attention from all channels, or Mathematics, which captures less social interest. Likewise, there are scientific fields where a certain channel is particularly relevant. We can conclude that a research dissemination plan or a transfer plan should be adapted to the area in which researchers publish. The reception of the scientific message has become a multidimensional phenomenon that we can trace in detail with altmetrics.

# Appendix

**Fig. 4** ESI categories and resume of the selection process form identification of top authors

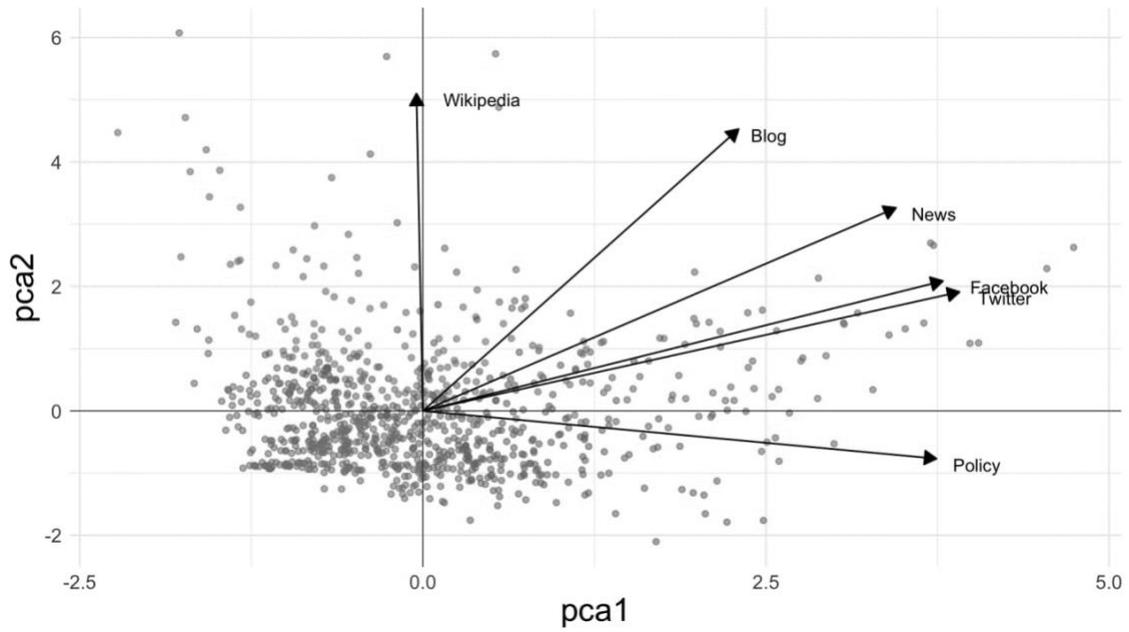

**Table 4** Principal Component Analysis for the 23 ESI fields

| ESI field | Twitter | | Facebook | | News | | Blogs | | Wikipedia | | Policy | |
|---|---|---|---|---|---|---|---|---|---|---|---|---|
| | pca | pca | pca | pca | pca | pca | pca | pca | pca | pca2 | pca | pca2 |
| Agricultural Sciences0 | 0,8 | | 0,8 | | 0,8 | | 0,7 | | | 0,82 | | 0,69 |
| Arts & Humanities | 0,8 | | 0,7 | | 0,8 | | 0,8 | | 0,3 | - | | 0,91 |
| Biology & Biochemistry | 0,8 | | 0,7 | | 0,8 | | 0,8 | | 0,5 | | | 0,99 |
| Chemistry | 0,7 | | 0,7 | | 0,7 | | 0,7 | | 0,3 | | | 0,96 |
| Clinical Medicine | 0,8 | | 0,7 | | 0,8 | | 0,8 | | | 0,92 | 0,3 | 0,51 |
| Computer Science | 0,8 | | 0,5 | | 0,3 | 0,5 | 0,7 | | | 0,78 | 0,4 | - |
| Economics & Business | 0,7 | | 0,5 | | 0,7 | | 0,7 | | | 0,96 | 0,3 | |
| Engineering | 0,6 | | 0,5 | | 0,7 | | 0,7 | | | | | 0,96 |
| Environment/Ecology | 0,6 | 0,5 | 0,5 | 0,6 | 0,7 | 0,4 | 0,6 | 0,5 | | 0,84 | 0,8 | |
| Geosciences | 0,8 | | 0,8 | | 0,8 | | 0,8 | | 0,7 | | | 0,97 |
| Immunology | 0,8 | | 0,7 | | 0,7 | | 0,7 | | | 0,87 | | 0,61 |
| Mathematics | 0,8 | | 0,6 | | 0,8 | | 0,7 | | 0,3 | | | 0,97 |
| Microbiology | 0,6 | 0,3 | | 0,6 | 0,7 | | 0,7 | | | 0,55 | | - |
| Molecular Biology & G. | 0,8 | | 0,7 | | 0,7 | | 0,7 | | 0,4 | - | | 0,83 |
| Material Science | 0,8 | | 0,8 | | 0,8 | | 0,8 | | 0,5 | | | 0,98 |
| Neuroscience & | 0,8 | | 0,7 | | 0,8 | | 0,8 | | | 0,58 | | 0,89 |
| Pharmacology & Tox. | 0,7 | | 0,7 | | 0,7 | | 0,7 | | | 0,74 | | 0,76 |
| Physics | 0,8 | | 0,7 | | 0,8 | | 0,8 | | 0,4 | | | 0,95 |
| Plant & Animal Science | 0,8 | | 0,8 | | 0,7 | | 0,7 | | 0,5 | | | 0,99 |
| Psychiatry/Psychology | 0,8 | | 0,7 | | 0,8 | | 0,8 | | | 0,64 | | 0,86 |
| Social Sciences. General | 0,7 | 0,3 | 0,7 | | 0,8 | | 0,8 | | 0,7 | | | 0,93 |
| Space Sciences | 0,8 | | 0,7 | | 0,8 | | 0,8 | | 0,7 | | | 0,99 |